\documentclass[aps,prl,superscriptaddress,preprintnumbers,twocolumn,groupedaddress,showpacs]{revtex4}
\usepackage{txfonts}
\usepackage{graphicx}

\begin{document}

\newcommand{\nb}{\nonumber}

\newcommand{\ppJW}{$pp \to J/\psi +W+X~$}

\title{ QCD corrections to $J/\psi$ production in association with a $W$-boson at the LHC    }

\author{ Li Gang$^{(a)}$, Song Mao$^{(b)}$, Zhang Ren-You$^{(b)}$, and Ma Wen-Gan$^{(b)}$   \\
{\small $^{(a)}$ School of Physics and Material Science, Anhui
University, Hefei, Anhui 230039, P.R.China} \\
{\small $^{(b)}$ Department of Modern Physics, University of Science and Technology of China (USTC), }  \\
{\small Hefei, Anhui 230026, P.R.China} }

\begin{abstract}
We calculate the next-to-leading order (NLO) QCD corrections to the
$J/\psi +W$ production at the LHC, and provide the theoretical
distribution of the $J/\psi$ transverse momentum. Our results show
that the differential cross section $\frac{d\sigma}{dp_T^{J/\psi}}$
at the LO is significantly enhanced by the NLO QCD
corrections. We believe that the comparison between the theoretical
predictions for the $J/\psi+W$ production and the
experimental data at the LHC can provide a verification for the colour-octet
mechanism of non-relativistic QCD in the
description of the processes involving heavy quarkonium.
\end{abstract}

\pacs{12.38.Bx, 12.39.St, 13.60.Le} \maketitle

\par
The study of heavy quarkonium is one of the most interesting
subjects in both theoretical and experimental physics, which offers
a good ground for investigating Quantum Chromodynamics (QCD)
in both perturbative and non-perturbative regimes. The factorization
formalism of nonrelativistic QCD (NRQCD) \cite{bbl} provides a
rigorous theoretical framework to describe the heavy-quarkonium
production and decay by separating the transition rate (production
cross section or decay rate) into two parts, the short-distance part
which can be expanded as a power series in $\alpha_s$ and calculated
perturbatively, and the long-distance matrix elements (LDMEs) which
can be extracted from experiments. The importance of the
LDMEs can be estimated by using velocity scaling rules
\cite{vrule}. A crucial feature of the NRQCD is that the complete
structure of the quarkonium Fock space has been explicitly
considered.

\par
By introducing the color-octet mechanism (COM), the NRQCD has
successfully absorbed the infrared divergences in P-wave
\cite{bbl,p1,p2} and D-wave \cite{d1,d2} decay widths of heavy
quarkonium, which can not be handled in the color-singlet mechanism
(CSM). The COM can successfully reconcile the orders of magnitude of
the discrepancies between the experimental data of $J/\psi$
production at the Tevatron \cite{comtev} and the CSM theoretical
predictions, even if they have been calculated up to the NLO. The
DELPHI data also favor the NRQCD COM predictions for the
$\gamma\gamma \to J/\psi+ X$ process \cite{DELPH1,DELPH2}. Similarly
the recent experimental data on the $J/\psi$ photoproduction of H1
\cite{H1} are fairly well described by the complete NLO NRQCD
corrections \cite{kniehl1}, and give a strong support to the
existence of the COM. However, the observed cross sections for
the double charmonium production at B factories
\cite{bf} are much larger than the LO NRQCD prediction \cite{nrbf}.
This discrepancy can be resolved by considering the CSM NLO
QCD corrections \cite{nloqcd} and the relativistic corrections
\cite{rel} without invoking the color-octet contributions
\cite{nocom}. Furthermore, the $J/\psi$ polarization in hadroproduction
at the Tevatron \cite{nnrtev} and photo-production at the HERA
\cite{nnrHERA} also conflict with the NRQCD predictions. Therefore, the
existence of the COM is still under doubt and far from being
proven. The further tests for the CSM and COM under the NRQCD
in heavy quarkonium production are still needed.

\par
In order to test the COM, it is an urgent task to
study the processes which significantly depend on the production
mechanism. The $J/\psi$ production associated with a $W$ boson
at the LHC, \ppJW, can serve as a such kind of process \cite{CCW}. For
this process, only the $^3S_1$ color-octet (the
$c\bar{c}[^3S_1^{(8)} ]$ Fock state) provides contribution at the
leading-order (LO). Even including the NLO QCD corrections up to
the $\alpha^3_s v^7$ order, there are only color-octets $c\bar{c} [
^1S_0^{(8)} ]$, $c\bar{c} [ ^3S_1^{(8)} ]$ and $c\bar{c} [
^3P_J^{(8)} ]$ $(J=0,1,2)$, but no color-singlet contribution exists in the
\ppJW process. Therefore, the $J/\psi+W$ production at the LHC is
an ideal ground to study the COM.

\par
\begin{figure}
\includegraphics[width=6cm]{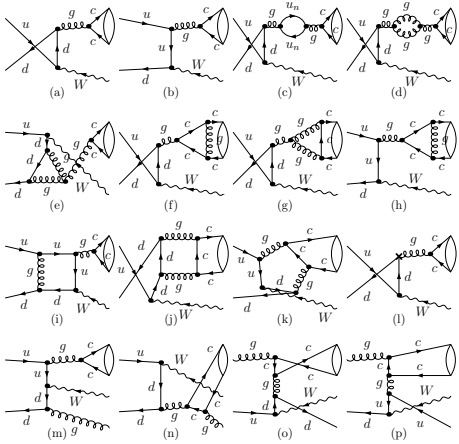}
\vspace*{-0.3cm} \caption{\label{fig1}  Representative Feynman
diagrams for $pp \to J/\psi+W^+$.}
\end{figure}

As we know, the NLO QCD corrections to quarkonium production are
usually significant \cite{nloqcd,nlo1,nlo2}. We should generally
take the NLO QCD corrections into account in studying the COM and
the universality of the LDMEs. In this paper, we calculate the
$J/\psi + W$ production at the LHC up to the $\alpha_s^3 v^7$
order within the NRQCD framework by applying the covariant
projection method \cite{p2}, and present the theoretical prediction
of the $p_T^{J/\psi}$ distribution. The LO cross section for the
parent process $pp \to J/\psi+W^{\pm}+X$ involves the contributions
of the following partonic processes,
\begin{eqnarray}
u\bar{d}\to c\bar{c}[^3S_1^{(8)}]+W^+, ~~ d\bar{u}\to
c\bar{c}[^3S_1^{(8)}]+W^-.
\end{eqnarray}
Since the cross sections for the $u\bar{d}\to
c\bar{c}[^3S_1^{(8)}]+W^+$ and $d\bar{u}\to
c\bar{c}[^3S_1^{(8)}]+W^-$ partonic processes are the same due to
the CP-conservation, we present only the detailed description
for the calculation of the partonic process $u(p_1)\bar{d}(p_2)\to
c\bar{c}[^3S_1^{(8)}](p_3)+W^+(p_4)$. The tree-level diagrams for
this partonic process are drawn in Figs.\ref{fig1}(a)-(b).

\par
In the nonrelativistic limit, by applying the covariant projection
method \cite{p2} we obtain the differential cross section
for $u\bar{d} \to c\bar{c}[^3S_1^{(8)}]+W^+$ expressed as
\begin{eqnarray}
&&\frac{d\hat{\sigma}_0}{d\hat{t}}=\frac{<{\cal
O}^{J/\psi}[^3S_1^{(8)}]>}{16\pi \hat{s}^2
N_{col}(^3S_1^{(8)})N_{pol}(^3S_1^{(8)})}\frac{64g^2{\alpha_s}^2
{\pi}^2}{9m_{J/\psi}^3\hat{t}^2\hat{u}^2}\times \nb\\
&\times& \left \{-m_w^2m_{J/\psi}^2
\left (\hat{t}^2 +\hat{u}^2\right )+\hat{t}\hat{u}\left [2\hat{s}^2+\hat{t}^2+
\hat{u}^2 \right. \right. \nb \\
&+& \left. \left. 2\hat{s}\left (\hat{t}+\hat{u}\right )\right ]\right \},
\end{eqnarray}
where $\hat{s}=(p_1+p_2)^2$, $\hat{t}=(p_1-p_3)^2$ and
$\hat{u}=(p_1-p_4)^2$. $N_{col}(^3S_1^{(8)})$ and
$N_{pol}(^3S_1^{(8)})$ refer to the color and polarization degrees
of freedom of $c\bar{c}[^3S_1^{(8)}]$ \cite{p2}, respectively.
Then the LO cross section for the $pp \to J/\psi+W^++X$ process is
\begin{eqnarray}
\sigma^{(0)} &=&\int dx_1dx_2d\hat{\sigma}_{0}
[G_{u/A}(x_1,\mu_f)G_{\bar{d}/B}(x_2,\mu_f)\nb\\&+&(A
\leftrightarrow B )],~~~~~~~~~~~~(\hat{s} = x_1 x_2 s),
\end{eqnarray}
where $G_{u,\bar d/A,B}$ are the parton distribution
functions (PDFs), and A, B represent the two incoming protons at the LHC.

\par
In calculating the NLO QCD corrections to the $pp \to J/\psi + W^+
+ X$ process, we should consider both the virtual correction and
the real gluon/light-quark emission correction. The virtual
corrections only come from the $c\bar{c} [
^3S_1^{(8)} ]$ Fock state, while the real gluon/light-quark
emission correction involves the contributions of the
$c\bar{c} [ ^1S_0^{(8)} ]$, $c\bar{c} [ ^3S_1^{(8)} ]$ and
$c\bar{c} [ ^3P_J^{(8)} ]$ $(J=0,1,2)$ Fock states.
In our calculations, we adopt the dimensional regularization
(DR) scheme to regularize the UV and IR
divergences, and the modified minimal subtraction ($\overline{{\rm
MS}}$) and on-mass-shell schemes to renormalize the strong
coupling constant and the quark wave functions, respectively.

\par
There are 41 virtual QCD one-loop diagrams for the subprocess
$u\bar{d}\to J/\psi+W^+$, which include self-energy (12), vertex
(10), box (7), pentagon (2) and counterterm (10) diagrams. We present part of
these diagrams in Figs.\ref{fig1}(c)-(l). There exist ultraviolet (UV),
Coulomb and soft/collinear infrared (IR) singularities in the
virtual correction. The UV singularities
are canceled by the counterterms of the strong coupling constant
and the quark wave functions after the renormalization procedure.
But the QCD one-loop amplitude of the partonic process
$u\bar{d}\to c\bar{c}[^3S_1^{(8)}]+W^+$ still contains Coulomb and
soft/collinear IR singularities. The IR and Coulomb singularities
in the virtual correction can be expressed as
\begin{eqnarray}
\label{virtual cross section}
d\hat{\sigma}^V=d\hat{\sigma}_{0}\left[\frac{\alpha_s}{2 \pi}
\frac{\Gamma(1-\epsilon)}{\Gamma(1-2 \epsilon)}\left(\frac{4 \pi
\mu_r^2}{\hat{s}}\right)^{\epsilon}\right]
\left(\frac{A_{2}^{V}}{\epsilon^2}+\frac{A_{1}^{V}}{\epsilon}-\frac{\pi^2}{6
v} +A_{0}^V\right),\nb \\
\end{eqnarray}
where
\begin{eqnarray}
A_{2}^{V}&=& -2 C_F,~~~~~~A_{1}^{V} = -7+3\left({\rm ln}\frac{4
m_c^2-\hat{t}}{m_c\sqrt{\hat{s}}}+{\rm ln}\frac{4
m_c^2-\hat{u}}{m_c\sqrt{\hat{s}}}\right),\nb\\C_F&=&4/3.
\end{eqnarray}

\par
The soft/collinear IR singularities can
be canceled by adding the contributions of the real gluon and
light-quark emission partonic processes, and redefining the parton
distribution functions at the NLO. For the Coulomb singularities,
they can be canceled after taking into account the corresponding corrections to
the operator $<{\cal O}^{J/\psi}[^3S_1^{(8)}]>$. We use
the expressions in Refs.\cite{OneTwoThree,Four,Five} to implement
the numerical evaluations of IR-safe one-point, 2-point, 3-point,
4-point and 5-point integrals. In our calculations, only two diagrams,
Figs.\ref{fig1}(f) and (h), contain the Coulomb singularities, which are
regularized by a small relative velocity $v$ between $c$ and
$\bar{c}$ \cite{coulomb}. We adopt the expressions in
Ref.\cite{IRDV} to deal with the IR-divergent Feynman
integral functions.

\par
The real gluon emission process provides three types of
corrections, which correspond to the contributions from the
$c\bar{c} [ ^1S_0^{(8)} ]$, $c\bar{c} [ ^3S_1^{(8)} ]$ and
$c\bar{c} [ ^3P_J^{(8)} ]$ $(J=0,1,2)$ Fock states, respectively.
The real gluon emission correction to the $u\bar{d}\to c\bar{c} [
^1S_0^{(8)} ]+W^+$ subprocess is free of divergence, and can be
numerically calculated by using the Monte Carlo method. For the
$c\bar{c} [ ^3S_1^{(8)} ]+W^+ + g$ production, it contains both
soft and collinear IR singularities which can be conveniently
isolated by adopting the two cutoff phase space slicing (TCPSS)
method \cite{TCPSS}.

\par
In adopting the TCPSS method, we should introduce two arbitrary
cutoffs, $\delta_s$ and $\delta_c$. The phase space of the
$u\bar{d}\to c\bar{c}[^3S_1^{(8)}]+W^++g$ subprocess can be split up
into two regions, $E_5 \leq \delta_s\sqrt{\hat{s}}/2 $ (soft gluon
region) and $E_5> \delta_s\sqrt{\hat{s}}/2 $ (hard gluon region) by
soft cutoff $\delta_s$. Furthermore, the hard gluon region is
separated as hard collinear (${\rm HC}$) and hard non-collinear
($\overline{\rm HC}$) regions. The ${\rm HC}$ region is the phase
space where  $-\hat{t}_{15}$ (or $-\hat{t}_{25}$)$ <\delta_c
\hat{s}$ $(\hat{t}_{15}\equiv(p_1-p_5)^2$ and
$\hat{t}_{25}\equiv(p_2-p_5)^2)$. Therefore, the cross section for
the real gluon emission subprocess can be expressed as
\begin{eqnarray}
\hat{\sigma}^R_g(^3S_1^{(8)})&=&\hat{\sigma}^S_g(^3S_1^{(8)})+\hat{\sigma}^H_g(^3S_1^{(8)})
\nb \\&=&\hat{\sigma}^S_g(^3S_1^{(8)})+\hat{\sigma}^{\rm
HC}_g(^3S_1^{(8)})+\hat{\sigma}^{\overline{\rm HC}}_g(^3S_1^{(8)}).\nb \\
\end{eqnarray}
The cross section for the subprocess $u\bar{d} \to
c\bar{c}[^3S_1^{(8)}]+W^++g$ in the hard non-collinear
($\overline{\rm HC}$) region is free of divergence, and can be
numerically calculated by using the Monte Carlo method. The
differential cross section for the subprocess $u\bar{d}\to
c\bar{c}[^3S_1^{(8)}]+W^++g$ in the soft region can be expressed
as
\begin{eqnarray}
\label{soft cross section} d\hat{\sigma}^S_g(^3S_1^{(8)})
&=&-\frac{\alpha_s}{2\pi}\left \{\frac{1}{6}\left[g(p_1,p_2)+g(p_c,p_{\bar{c}})\right]
\right. \nb\\
&& \left.-\frac{7}{6}\left[g(p_1,p_c)+g(p_2,p_{\bar{c}})\right]
-\frac{1}{3}\left[g(p_1,p_{\bar{c}})\right. \right.\nb\\
&& \left.\left.+g(p_2,p_c)\right]\right\}d\hat{\sigma}_{0},
\end{eqnarray}
where $g(p_i,p_j)$ are soft integral functions defined as \cite{c2
been,c2 willy,c2 hs}
\begin{eqnarray}
g(p_i,p_j)&=&\frac{(2\pi\mu_r)^{2\epsilon}}{2\pi}\int_{E_5\leq\delta_s\sqrt{\hat{s}}/2}
\frac{d^{D-1}p_5}{E_5}\left[\frac{2(p_i \cdot p_j)}{(p_i\cdot
p_5)(p_j\cdot p_5)}
\right.\nb\\
&&\left. -\frac{p^2_i}{(p_i \cdot p_5)^2}-\frac{p^2_j}{(p_j\cdot
p_5)^2}\right].
\end{eqnarray}
Then we can get
\begin{eqnarray}
\label{soft cross section1} d\hat{\sigma}^S_g(^3S_1^{(8)})
&=&d\hat{\sigma}_{0} \left[\frac{\alpha_s}{2 \pi}
\frac{\Gamma(1-\epsilon)}{\Gamma(1-2 \epsilon)}\left(\frac{4 \pi
\mu_r^2}{\hat{s}}\right)^{\epsilon}\right] \times \nb \\
&&
\times\left(\frac{A_{2}^{S}}{\epsilon^2}+\frac{A_{1}^{S}}{\epsilon}
+A_{0}^S\right),
\end{eqnarray}
where
\begin{eqnarray}
A_{2}^{S}&=& 2 C_F,~~~~A_{1}^{S} = 3-3\left({\rm ln}\frac{4 m_c^2-\hat{t}}
{m_c\sqrt{\hat{s}}}+{\rm ln}\frac{4 m_c^2-\hat{u}}{m_c\sqrt{\hat{s}}}\right).\nb\\
\end{eqnarray}
For the subprocess $u\bar{d}\to c\bar{c}[^3S_1^{(8)}]+W^++g$ only
the gluon radiation from initial particles can induce collinear
singularities. The differential cross section,
$d\hat{\sigma}^{HC}_{g}$, can be written as
\begin{eqnarray}\nb
d\sigma^{HC}_g&=&d\hat{\sigma}_{0} \left[\frac{\alpha_s}{2\pi}
\frac{\Gamma(1-\epsilon)}{\Gamma(1-2\epsilon)}\left(\frac{4\pi\mu^2_r}{\hat
s}\right)^\epsilon \right]\left(-\frac{1}{\epsilon}\right)\times \nb\\
&\times &\delta_c^{-\epsilon}
\left [P_{uu}(z,\epsilon)G_{u/A}(x_1/z,\mu_f)G_{\bar d /B}(x_2,\mu_f) +
\right. \nb\\
&\times & \left. P_{\bar d \bar d}(z,\epsilon) G_{\bar
d/B}(x_2/z,\mu_f)G_{u/A}(x_1,\mu_f)+(A\leftrightarrow B)
\right ] \nb \\
&\times & \frac{dz}{z}\left(\frac{1-z}{z}\right)^{-\epsilon}dx_1dx_2,
\label{HCi}
\end{eqnarray}
where $P_{uu}(z,\epsilon)$ and $P_{\bar d\bar d}(z,\epsilon)$ are
the D-dimensional unregulated ($z <1$) splitting functions related
to the usual Altarelli-Parisi splitting kernels \cite{APsk}.
$P_{ii}(z,\epsilon) (i=u, \bar d)$ can be written explicitly as
\begin{eqnarray}\nb
\label{Peq0} P_{ii}(z,\epsilon)&=& P_{ii}(z) + \epsilon
P'_{ii}(z)\\\nb
P_{ii}(z) &=& C_F \frac{1+z^2}{1-z}\\
P'_{ii}(z) &=& -C_F(1-z)~~~~~(i=u, \bar d).   \label{Pii}
\end{eqnarray}

\par
As for the partonic process $u\bar{d}\to
c\bar{c}[^3P_J^{(8)}]+W^++g$, it contains only the soft
singularities. Using the TCPSS method mentioned above, we split the
phase space up into soft gluon region and hard gluon region by
adopting the cutoff $\delta_s$. Then the cross section for the
partonic process $u\bar{d}\to c\bar{c}[^3P_J^{(8)}]+W^++g$ can be
expressed as
\begin{eqnarray}
\hat{\sigma}^R_g(^3P_J^{(8)}) =\hat{\sigma}^{S}_g(^3P_J^{(8)}) +
\hat{\sigma}^{H}_g(^3P_J^{(8)}).
\end{eqnarray}
The cross section $\hat{\sigma}^{H}_g(^3P_J^{(8)})$ is finite and
can be evaluated in four dimensions by using Monte Carlo method. The
differential cross section in the soft region for the process
$u\bar{d}\to c\bar{c}[^3P_J^{(8)}]+W^++g$,
$d\hat{\sigma}^{S}_g(^3P_J^{(8)})$, can be expressed as
\begin{eqnarray}
\label{collinear-d}
d\hat{\sigma}^{S}_g(^3P_J^{(8)})&=&-\left(\frac{1}{\epsilon}-2{\rm
ln}\delta_s+\frac{1}{\beta}{\rm
ln}\frac{1+\beta}{1-\beta}\right)\frac{4 \alpha_s B_F}{3\pi
m^2_c}\nb \\
&\times&\frac{\Gamma(1-\epsilon)}{\Gamma(1-2 \epsilon)}\left(\frac{4
\pi
\mu_r^2}{\hat{s}}\right)^{\epsilon} <{\cal O}^{J/\psi}\left[{}^3\!P_J^{(8)}\right]> \nb\\
&\times&\frac{d\hat{\sigma}_0}{<{\cal
O}^{J/\psi}[^3S_1^{(8)}]>} \nb\\
\end{eqnarray}
with $\beta=\sqrt{1-4 m_c^2/E_3^2}$ and $E_3=\frac{\hat{s}+4
m_c^2-m_w^2}{2 \sqrt{\hat{s}}}$.

\par
The real light-quark corrections to the subprocess $u\bar{d}\to
J/\psi+W^+$ arise from the partonic processes
\begin{eqnarray} g(p_1)u(\bar{d})(p_2)\to
c\bar{c}[n](p_3)+W^+(p_4)+d(\bar{u})(p_5),
\end{eqnarray}
where $n=~ ^3S_1^{(8)}$, $^1S_0^{(8)}$ and $^3P_J^{(8)}$. Some of the
Feynman diagrams for these partonic processes are presented in
Figs.\ref{fig1}(o)-(p).

\par
The real light-quark partonic processes with $n=~^1S_0^{(8)}$ and
$^3P_J^{(8)}$ contain no singularities, so we can perform their
phase space integrations by using the general Monte Carlo method.
The partonic processes $gu(\bar{d})\to
c\bar{c}[^3S_1^{(8)}]+W^++d(\bar{u})$ contain only the initial state
collinear singularities. By using the TCPSS method, we split the
phase space up into two regions, collinear region and non-collinear
region,
\begin{eqnarray}
\hat{\sigma}^R_{q}(^3S_1^{(8)}) = \hat{\sigma}^{\rm
C}_{q}(^3S_1^{(8)}) +
\hat{\sigma}^{\overline{C}}_{q}(^3S_1^{(8)})~~~~ (q= u, \bar{d}).
\end{eqnarray}
The cross section in non-collinear region,
$\hat{\sigma}_q^{\overline{C}}(^3 S_1^{(8)})$ is finite and can be
evaluated in four dimensions by using the Monte Carlo method. The
differential cross sections for the subprocesses $gu(\bar{d})\to
c\bar{c}(^3S_1^{(8)})+W^++d(\bar{u})$ can be written as
\begin{eqnarray}
\label{collinear-b} d\sigma^{C}_q (^3S_1^{(8)})&=& d\hat{\sigma}_{0}
\left[\frac{\alpha_s}{2 \pi} \frac{\Gamma(1-\epsilon)}{\Gamma(1-2
\epsilon)}\left(\frac{4 \pi
\mu_r^2}{\hat{s}}\right)^{\epsilon}\right]\left(-\frac{1}{\epsilon} \right) \nb\\
&\times& \delta_c^{-\epsilon} [ P_{\bar{d}(
u)g}(z,\epsilon)G_{g/A}(x_1/z,\mu_f)G_{u(\bar d)/B}(x_2,\mu_f)\nb\\
&+&(A\leftrightarrow B)]\frac{dz}{z}\left( \frac{1-z}{z}
\right)^{-\epsilon }dx_1dx_2
\end{eqnarray}
with $q = u, \bar{d}$ and $P_{qg}(z,\epsilon)$ can be expressed explicitly as
\begin{eqnarray}
\label{Peq1}
&&P_{qg}(z,\epsilon)=P_{qg}(z)+ \epsilon P'_{qg}(z), \nb \\
&&P_{qg}(z)=\frac{1}{2}[z^2+(1-z)^2],\nb \\&&P'_{qg}(z) = -z(1-z).
\end{eqnarray}

\par
To obtain an IR-safe cross section for the $pp\to J/\psi + W^++X$ up
to the NLO, we should take into account both the NLO QCD counterterms of
the PDFs and the NLO QCD corrections to the operator $<{\cal
O}^{J/\psi}[^3S_1^{(8)}]>$. The ${\cal O}(\alpha_s)$ counterterms of
the PDFs are expressed as
\cite{Harris}
\begin{eqnarray}
\delta G_{i/A}(x,\mu_f)&=& \frac{1}{\epsilon}
\left[\frac{\alpha_s}{2 \pi} \frac{\Gamma(1-\epsilon)}{\Gamma(1-2
\epsilon)}\left(\frac{4 \pi
\mu_r^2}{\mu_f^2}\right)^{\epsilon}\right]\nb\\
&\times & \int^1_z\frac{dz}{z}P_{ij}(z)G_{j/A}(x/z,\mu_f).
\end{eqnarray}
By adding the contributions of the PDF counterterms and the real
gluon/light-quark emission collinear corrections shown in
Eqs.(\ref{HCi}) and (\ref{collinear-b}),we obtain
\begin{eqnarray}
\label{collinear cross section1}
 d\sigma^{coll}&=&d\hat{\sigma}_{0}
\left[\frac{\alpha_s}{2 \pi} \frac{\Gamma(1-\epsilon)}{\Gamma(1-2
\epsilon)}\left(\frac{4 \pi
\mu_r^2}{\hat{s}}\right)^{\epsilon}\right] \left \{
\tilde{G}_{u/A}(x_1,\mu_f)\right. \nb\\
&&\left.  \times
G_{\bar{d}/B}(x_2,\mu_f)+G_{u/A}(x_1,\mu_f)\tilde{G}_{\bar{d}/B}(x_2,\mu_f)
\right.\nb \\
&+&\left. \sum_{\alpha=u,\bar{d}} \left[\frac{A_1^{sc}(\alpha \to
\alpha g)}{\epsilon}+A_0^{sc}(\alpha \to \alpha g)
\right]G_{u/A}(x_1,\mu_f)\right. \nb\\
&&\left.  \times G_{\bar{d}/B}(x_2,\mu_f) + (A\leftrightarrow
B)\right\} dx_1dx_2,
\end{eqnarray}
where
\begin{eqnarray}
&&A_1^{sc}(\alpha \to \alpha g)= C_F(2 \ln \delta_s+3/2),\nb
\\
&&A_0^{sc} = A_1^{sc} \ln\left (\frac{\hat{s}}{\mu_f^2}\right
),~~~~~~~~~~\alpha=u,\bar{d}
\end{eqnarray}
and
\begin{eqnarray}
&&\tilde{G}_{\alpha/H}(x,\mu_f) =
\sum_{\alpha^{\prime}=\alpha,g}\int^{1-\delta_s \delta_{\alpha
\alpha^{\prime}}}_x \frac{dy}{y} G_{\alpha^{\prime}/H}(x/y,\mu_f)
\tilde{P}_{\alpha
\alpha^{\prime}}(y),\nb\\&&(H=A,B,~\alpha=u,\bar{d}),
\end{eqnarray}
with
\begin{eqnarray}
\tilde{P}_{\alpha \alpha^{\prime}}(y) = P_{\alpha \alpha^{\prime}}
\ln \left( \delta_c \frac{1-y}{y} \frac{\hat{s}}{\mu_f^2} \right) -
P^{\prime}_{\alpha \alpha^{\prime}}(y).
\end{eqnarray}

\par
\begin{figure}
\includegraphics[width=8cm]{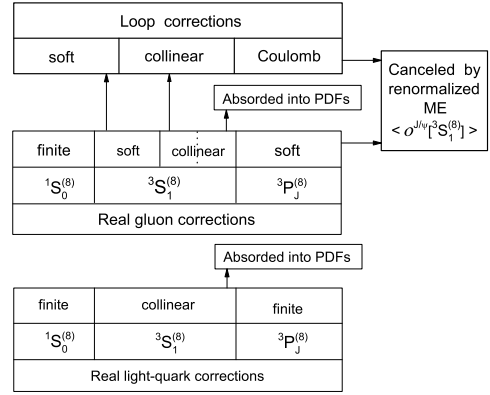}
\vspace*{-0.3cm} \caption{\label{fig2}  The IR and Coulomb
singularities structure in the NLO QCD calculations for the $pp
\to J/\psi + W^+ + X$ process. }
\end{figure}

We can see that the summation of the soft (Eq.(\ref{soft cross
section1})), collinear (Eq.(\ref{collinear cross section1})), and UV
renormalized virtual corrections (Eq.(\ref{virtual cross section}))
to the $pp \to J/\psi + W^+ + X$ process, $d\sigma^{S}_g+
d\sigma^{coll} + d\sigma^{V}$, is soft/collinear IR- and UV-finite,
i.e.,
\begin{eqnarray}
A^S_2&+&A^V_2=0, \nb \\
A^S_1&+&A^V_1+ A_1^{sc}(u\to ug)+A_1^{sc}(\bar d\to \bar dg)=0.\nb\\
\end{eqnarray}
However the above result, $d\sigma^{S}_g+ d\sigma^{coll} +
d\sigma^{V}$, still contains the Coulomb singularity. Furthermore,
the corrections contributed by the $c\bar{c} [^3P_J^{(8)} ]$
$(J=0,1,2)$ Fock states, $d \hat{\sigma}^R_g(^3P_J^{(8)})$, contain
soft IR singularities too. We can eliminate these singularities by
taking into account the NLO QCD corrections to the operator $<{\cal
O}^{J/\psi}[^3S_1^{(8)}]>$. In this paper we use the method in
Ref.\cite{p2} to deal with these singularities. In Fig.\ref{fig2} we
present the IR and Coulomb singularities structure in the NLO QCD
calculations for the $pp \to J/\psi + W^+ + X$ process. We have
checked analytically that all the IR and Coulomb singularities are
canceled in the final result.

\par
The final result for the process $pp\to J/\psi+W^+X$ up to the NLO
consists of three parts of contributions,
\begin{eqnarray}
\sigma_{total}=\sigma_{^3S_1^{(8)}}+\sigma_{^1S_0^{(8)}}+\sigma_{^3P_J^{(8)}},
\end{eqnarray}
where $\sigma_{^3S_1^{(8)}}$ can be divided into two parts: a
two-body term $\sigma^{(2)}_{^3S_1^{(8)}}$ and a three-body term
$\sigma^{(3)}_{^3S_1^{(8)}}$. The two-body term,
$\sigma^{(2)}_{^3S_1^{(8)}}$, is expressed as
\begin{eqnarray}
\sigma^{(2)}_{^3S_1^{(8)}}&=& \sigma^{(0)}+\frac{\alpha_s}{2 \pi}
\int dx_1dx_2 d\hat{\sigma}_{0} \{ G_{u/A}(x_1,\mu_f)\nb\\&\times
& G_{\bar{d}/B}(x_2,\mu_f)[A^S_0+A^V_0+A_0^{sc}(u\to ug)\nb \\
&+&A_0^{sc}(\bar d\to \bar dg)] + \tilde{G}_{u/A}(x_1,\mu_f)G_{\bar
d/B}(x_2,\mu_f) \nb \\
&+&G_{u/A}(x_1,\mu_f)\tilde{G}_{\bar d/B}(x_2,\mu_f)+(A
\leftrightarrow B ) \}.
\end{eqnarray}
And the three-body term, $\sigma^{(3)}_{^3S_1^{(8)}}$, is
written as
\begin{eqnarray}
\sigma^{(3)}_{^3S_1^{(8)}}=\sigma^{\overline{\rm HC}}_g(^3S_1^{(8)})
+\sigma^{\overline{\rm C}}_q(^3S_1^{(8)}).
\end{eqnarray}

\par
$\sigma_{^3P_J^{(8)}}$ also can be divided into
two parts: a two-body term $\sigma^{(2)}_{^3P_J^{(8)}}$ and a
three-body term $\sigma^{(3)}_{^3P_J^{(8)}}$. The two-body
term, $\sigma^{(2)}_{^3P_J^{(8)}}$, is expressed as
\begin{eqnarray}
\sigma^{(2)}_{^3P_J^{(8)}}&=&\int dx_1dx_2
d\hat{\sigma}^{S}_g(^3P_J^{(8)})\{
G_{u/A}(x_1,\mu_f)G_{\bar{d}/B}(x_2,\mu_f)\nb\\&+&(A \leftrightarrow
B ) \}.
\end{eqnarray}
And the three-body term,
$\sigma^{(3)}_{^3P_J^{(8)}}$, is written as
\begin{eqnarray}
\sigma^{(3)}_{^3P_J^{(8)}}=\sigma^{H}_g(^3P_J^{(8)})
+\sigma_q(^3P_J^{(8)}).
\end{eqnarray}
Finally, after taking into account the NRQCD NLO corrections to the
operator $<{\cal O}^{J/\psi}[^3S_1^{(8)}]>$, the contributions of
the $^3S_1^{(8)}$ and $^3P_J^{(8)}$ states are finite. As for the
contribution from $^1S_0^{(8)}$ state, it contains no singularity
and only involves a three-body term $\sigma^{(3)}_{^1S_0^{(8)}}$,
which can be expressed as
\begin{eqnarray}
\sigma_{^1S_0^{(8)}}= \sigma_g(^1S_0^{(8)}) +\sigma_q(^1S_0^{(8)}).
\end{eqnarray}

\par
As a check of the correctness of our calculations, the independence
of the cross section part of
$\sigma_{^3S_1^{(8)}}+\sigma_{^1S_0^{(8)}} +\sigma_{^3P_J^{(8)}}$,
on the two arbitrary cutoffs, $\delta_s$ and $\delta_c$, has been
numerically verified.

\par
For the $pp\to J/\psi+W+X$ process at the LHC, we take CTEQ6L1 PDFs
with an one-loop running $\alpha_s$ in the LO calculations, and
CTEQ6M PDFs with a two-loop $\alpha_s$ in the NLO calculations
\cite{CTEQ6}. For the QCD parameters we take the number of active
flavor as $n_f=3$, and input $\Lambda_{{\rm QCD}}^{(3)} = 249~{\rm
MeV}$ for the LO and $\Lambda_{{\rm QCD}}^{(3)} = 389~{\rm MeV}$ for
the NLO calculations \cite{kniehl1}, respectively. The masses of the
external particles and the fine structure constant are taken as $m_W
= 80.398~{\rm GeV}$, $m_c = m_{J/\psi}/2 = 1.5~{\rm GeV}$ and
$\alpha = 1/137.036$. The sine squared of the Weinberg angle is
expressed as $s_W^2=1-m_W^2/m_Z^2$, where $m_Z = 91.1876~{\rm GeV}$.
The renormalization, factorization and NRQCD scales are chosen as
$\mu_r = \mu_f = m_T$ and $\mu_{\Lambda} = m_c$, respectively, where
$m_T = \sqrt{\Big( p_T^{J/\psi} \Big)^2 + m_{J/\psi}^2}$ is the
$J/\psi$ transverse mass. Since the $< {\cal O}^{J/\psi} [
^3P_J^{(8)} ]>$ $(J=0,1,2)$ LDMEs satisfy the multiplicity relations
\begin{eqnarray}
< {\cal O}^{J/\psi} [ ^3P_J^{(8)} ]> = (2J+1) < {\cal O}^{J/\psi} [
^3P_0^{(8)} ]>, \nonumber
\end{eqnarray}
we adopt the LDME $< {\cal O}^{J/\psi} [ ^3S_1^{(8)}]>=2.73 \times
10^{-3}~{\rm GeV}^3$ and the linear combination
\begin{eqnarray}
M_r^{J/\psi} = < {\cal O}^{J/\psi} [ ^1S_0^{(8)} ]> +
\frac{r}{m_c^2} < {\cal O}^{J/\psi} [ ^3P_0^{(8)} ]> \nonumber
\end{eqnarray}
with $M_r^{J/\psi}=5.72 \times 10^{-3}~{\rm GeV}^3$ and $r = 3.54$
as the input parameters, which were fitted to the Tevatron RUN-I
data by using the CTEQ4 PDFs and have taken into account the
dominant higher-order effects due to the multiple-gluon radiation in
the inclusive $J/\psi$ hadroproduction \cite{Kniehl:1998qy}. Then
the $< {\cal O}^{J/\psi} [ ^1S_0^{(8)} ]>$ and $< {\cal O}^{J/\psi}
[ ^3P_0^{(8)} ]>$ LDMEs are fixed by the democratic choice
\cite{Klasen:2004tz}
\begin{eqnarray}
< {\cal O}^{J/\psi} [ ^1S_0^{(8)} ]> = \frac{r}{m_c^2} < {\cal
O}^{J/\psi} [ ^3P_0^{(8)} ]> = \frac{1}{2} M_r^{J/\psi}. \nonumber
\end{eqnarray}

\par
In the calculations of the real corrections, the two phase space
cutoffs are chosen as $\delta_s = 10^{-3}$ and $\delta_c =
\delta_s/50$, and the invariance of the $\delta_s$ value running in
the range of $10^{-4}-10^{-2}$ is checked within the error
tolerance. Considering the validity of the NRQCD and perturbation
method, we restrict our results in the range of $p_T^{J/\psi} >
3~{\rm GeV}$ and $|y_{J/\psi}| < 3$.

\par
\begin{figure}
\includegraphics[width=9cm]{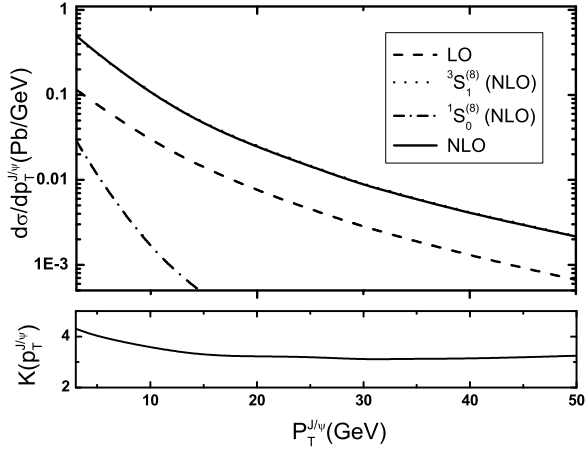}
\vspace*{-0.3cm}\caption{\label{fig3} The LO and NLO QCD corrected
distributions of $p_T^{J/\psi}$ for the $pp \to J/\psi+W^{\pm}+X$
process, and the contributions of the $c\bar{c} \left[ ^1S_0^{(8)}
\right]$ and $c\bar{c} \left[ ^3S_1^{(8)} \right]$ Fock states up to
NLO at the LHC.}
\end{figure}
\begin{figure}
\includegraphics[width=9cm]{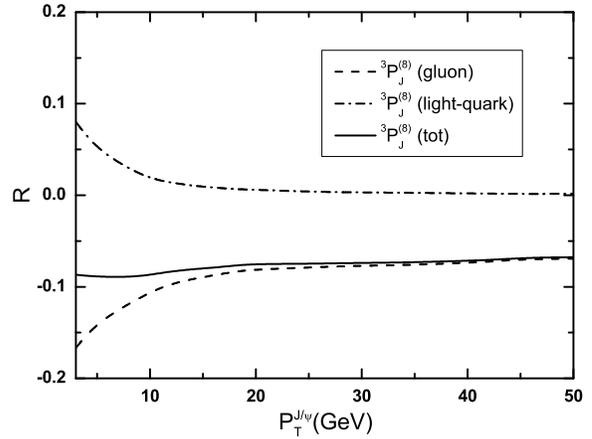}
\vspace*{-0.3cm}\caption{\label{fig4} The ratio $R$ versus
$p_T^{J/\psi}$ with the definition of $R \equiv
\frac{d\sigma^{^3P_J^{(8)}}}{d p_T^{J/\psi}}/\frac{d\sigma^{LO}}{d
p_T^{J/\psi}}$. }
\end{figure}

In Fig.\ref{fig3}, we present the LO and NLO QCD corrected
distributions of $p_T^{J/\psi}$ for the $pp\to J/\psi+W^{\pm}+X$
process at the LHC. For comparison, we also depict the contributions
of the $c\bar{c} [ ^1S_0^{(8)} ]$ and $c\bar{c} [ ^3S_1^{(8)} ]$
Fock states in the figure, while the total contribution of the
$c\bar{c} [ ^3P_J^{(8)} ]$ ($J=0,1,2$) Fock states is negative and
will be drawn in Fig.\ref{fig4}. From the Fig.\ref{fig3} we can see
that the differential cross section at the LO is significantly
enhanced by the QCD corrections. In the range of $3~{\rm GeV} <
p_T^{J/\psi} < 50~{\rm GeV}$, the $K$-factor, defined as $K =
\frac{d\sigma^{NLO}}{dp_T^{J/\psi}}/\frac{d\sigma^{LO}}{dp_T^{J/\psi}}$,
is in the range of $[3.09,~4.31]$, and reaches its maximum when
$p_T^{J/\psi} = 3~{\rm GeV}$. In Fig.\ref{fig4}, we present the
various contributions of the $c\bar{c} [ ^3P_J^{(8)} ]$ Fock states.
The solid, dashed and dash-dotted curves represent the total
contribution of the $c\bar{c} [ ^3P_J^{(8)} ]$ Fock states, the real
gluon emission correction to the $c\bar{c} [ ^3P_J^{(8)}] + W$
production and the real light-quark emission correction to the
$c\bar{c} [ ^3P_J^{(8)} ] + W$ production, respectively. In
Figs.\ref{fig3} and \ref{fig4}, we can see that there exists a
compensation between the contributions of the $S$ and $P$ states.
The contributions of the $c\bar{c} [ ^1S_0^{(8)} ]$, $c\bar{c} [
^3S_1^{(8)}]$ Fock states and the real light-quark emission
correction to the $c\bar{c} [ ^3P_J^{(8)} ] + W$ production are
always positive, while the real gluon emission correction to the
$c\bar{c} [ ^3P_J^{(8)} ] + W$ production is negative.

\par
In above calculations, we mainly consider the direct production of
$J/\psi$ mesons up to the NLO. But the $J/\psi$ also can be
produced indirectly via radiative or hadronic decays of heavier
charmonia, such as $\chi_{cJ}$ and $\psi^\prime$ mesons. The
respective decay branching fractions are $B(\chi_{c0}\to
J/\psi+\gamma)=(1.28\pm0.11)\%$, $B(\chi_{c1}\to
J/\psi+\gamma)=(36.0\pm1.9)\%$, $B(\chi_{c2}\to
J/\psi+\gamma)=(20.0\pm1.0)\%$ and $B(\psi^\prime\to
J/\psi+X)=(57.4\pm0.9)\%$ \cite{pdg}. The cross sections for these
four indirect production channels can be obtained approximately by multiplying
the direct-production cross sections for the respective intermediate
charmonia by their decay branching fractions.
At the LO, only the $c\bar{c} [ ^3S_1^{(8)} ]$
color-octet contributes to the productions of the charmonia
associated with a $W$ boson at the LHC. With the multiplicity
relations of LDMEs $<{\cal O}^{\chi_{cJ}}[{}^3\!S_1^{(8)}]>$,
$<{\cal O}^{\chi_{cJ}}[{}^3\!S_1^{(8)}]> =(2J+1) <{\cal
O}^{\chi_{c0}}[{}^3\!S_1^{(8)}]>$,  the cross sections for
the productions of the charmonia associated with $W$ can be expressed as
\begin{eqnarray}
d\hat{\sigma}(pp\to H+W)=d\hat{\sigma}_{0}\frac{<{\cal
O}^H[^3S_1^{(8)}]>}{<{\cal
O}^{J/\psi}[^3S_1^{(8)}]>},\nb\\
\end{eqnarray}
where $H$ represents the intermediate charmonia $\psi^\prime $ or
$\chi_{cJ}$. In numerical calculations, we adopt the LDMEs $<{\cal
O}^{\chi_{c0}}[{}^3\!S_1^{(8)}]>=(6.81\pm 1.75) \times 10^{-4}~{\rm
GeV}^3$\cite{Kniehl:1998qy} and $<{\cal
O}^{\psi^\prime}[{}^3\!S_1^{(8)}]>=(2.0\pm 0.2) \times 10^{-3}~{\rm
GeV}^3$\cite{Kniehl:31}. We find that the cross section for the
indirect $J/\psi$ production is almost the same as the LO cross
section for the direct $J/\psi$ production, $\sigma^{{\rm indirect}}
= 0.94 \times \sigma^{(0)}$. The complete indirect $J/\psi$
production in association with a W gauge boson at the NLO will be
calculated in our further work.

\par
In this paper we investigate the NLO QCD corrections to the $J/\psi
+W$ production at the LHC, and present the numerical results of the
differential cross section of the $p_T^{J/\psi}$ and the
contributions of the different Fock states for the \ppJW process up
to the QCD NLO. We find that the differential cross section at the
LO is significantly enhanced by the QCD corrections. The numerical
results show that there exists a negative real gluon emission
correction to the $c\bar{c} [ ^3P_J^{(8)} ] + W$ production. The LO
differential cross section for the process $pp \to J/\psi+W^{\pm}+X$
in the low $p_T^{J/\psi}$ region is heavily enhanced, and the NLO
QCD corrected differential cross section can reach $0.49~pb$ in the
vicinity of $p_T^{J/\psi} \sim 3~GeV$. Although the $J/\psi$ events
are difficult to be accepted at low $p_T$ region, we can find that
the cross section for $J/\psi +W$ direct production at the NLO is
about 0.81 $pb$ with $p_T^{J/\psi} > 10 GeV$ . To obtain the cross
section for the process $pp \to J/\psi +W+X$ with the pure leptonic
decays, we multiply the cross section for the direct production by
the branching fractions $12 \% $ for $J/\psi \to l^{+} l^{-}$ and
$22\%$ for $W \to l \nu$. Given the integrated luminosity of 300
$fb^{-1}$ at the LHC, we could obtain about 6400 events. If we
include the indirect contribution of the radiative or hadronic
decays of $\chi_{cJ}$ and $\psi^\prime$ mesons to $J/\psi$, more
events could be detected. Even taking into account the detector
acceptance and efficiency, there are still enough $J/\psi +W$ events
which can be detected. We conclude that the LHC has the potential to
detect the $J/\psi+W$ production. If the $J/\psi+W$ production is
really detected, it would be a solid basis for testing the
colour-octet mechanism of the NRQCD.

\vskip 10mm
\par
\noindent{\large\bf Acknowledgments:} This work was supported in
part by the National Natural Science Foundation of
China(No.10875112, No.11075150, No.11005101), the Specialized
Research Fund for the Doctoral Program of Higher
Education(No.20093402110030), and the 211 Project of Anhui
University.

\end{document}